\documentclass[12pt]{iopart}

\expandafter\let\csname equation*\endcsname\relax

\expandafter\let\csname endequation*\endcsname\relax

\usepackage{amsmath}
\usepackage{iopams}
\usepackage{amsfonts}
\usepackage{amssymb}
\usepackage{graphicx}

\begin{document}

\title[Bound states in the continuum and exceptional points in dielectric waveguide]{Bound states in the continuum and exceptional points in dielectric waveguide equipped with a metal grating}

\author{Ryo Kikkawa$^1$, Munehiro Nishida$^1$, and Yutaka Kadoya$^1$}

\address{$^1$Graduate School of Advanced Sciences of Matter, Hiroshima University, 1-3-1 Kagamiyama, Higashihiroshima, Hiroshima 7398530, Japan}

\ead{mnishida@hiroshima-u.ac.jp}

\begin{abstract}
Bound states in the continuum (BICs) and exceptional points (EPs) have been the subjects of recent intensive research as they exhibit exotic phenomena that are significant  for both fundamental physics and practical applications. 
We investigated the emergence of the Friedrich-Wintgen (FW) type BIC and the EP in a dielectric waveguide comprising a metal grating, focusing on their dependence on the grating thickness.
The BIC emerges at a branch near the anti-crossing formed of the two waveguide modes, for a grating of any thickness.
With the grating-thickness change, the anti-crossing gap varies and the branch at which the BIC appears flips.
We show that, when the slit is single mode, the BIC appears at the crossing between the two waveguide modes in the empty-lattice (zero slit-width) limit, while the results satisfy the criteria for the branch at which the BIC appears in the previous reports.
In addition, we find that the EP appears near the BIC in the same device only on selecting the grating thickness.
The BIC and EP in the dielectric waveguide comprising a metal grating, particularly with such tunability, are expected to result in the development of functional and high-performance photonic devices in addition to being a platform for the fundamental research of non-Hermitian systems.
\end{abstract}

\noindent{\it Keywords\/}: bound states in the continuum, dielectric waveguide, grating

\maketitle

\section{Introduction \label{Introduction}}

Resonant oscillations in open resonator systems (the system that possesses the radiation port to the far region) normally decay with time because of the radiative dissipation of the energy.
Even in open systems, however, purely bound states can exist when some requisite conditions are fulfilled.
Such a non-decaying state in an open system is called a bound state in the continuum (BIC)\cite{rev_nat, rev_nano} and was originally discussed in quantum mechanics in 1929 by von Neumann and Wigner\cite{1st_bic}.
In optics, BIC was first studied theoretically in 2008 \cite{bic2008} and then experimentally in a waveguide array in 2011\cite{bic2011}.
As BIC enables strong confinement of light, which results in high-Q resonances\cite{Nat2013_499_188}, applications such as single mode lasers\cite{biclaser1, biclaser2}, high-efficiency harmonic generation\cite{bicnonleniar1, bicnonleniar2, bicnonleniar3}, and biosensors\cite{bicsensor} have been demonstrated.
BICs can be categorized into several types based on their physical origin\cite{rev_nat}.
Among them, two types of BIC, the symmetry-protected BIC and accidental or Friedrich-Wintgen (FW) BIC\cite{bicfw}, have been mostly investigated.
The former originates from the incompatibility of the symmetry between the resonant mode and external radiation and appears at highly symmetric points such as the $\Gamma$ point in the reciprocal space of the periodic structure.
The latter is formed by the destructive interference between the radiation from the resonant modes\cite{rev_nat, SciRep2016_6_31908}, where the symmetry is generally not required, and can appear at a point of no symmetry in the reciprocal space.

When the two relevant modes interact in the near field, the dispersion of the modes anti-crosses, and the BIC lies on one of the split branches\cite{bicfw,PRB2006_73_235342,PRL2006_97_253901}.
In optics, recently, BICs appearing on the anti-crossed branch have been attracting attention and are being discussed\cite{bicmicro1, bicmicro2, bicatphot2,bicathyb} for applications such as high-Q super cavities with subwavelength dielectric structures\cite{bicmicro2}.
In our previous report\cite{KikkawaNJP}, we presented that, in a dielectric waveguide connected with the far field through a metal grating, the branch at which the BIC appears depends on the polarization of the mode, which is explained by the difference in the parameter describing the coupling phase of the modes with the external radiation, in a consistent manner with the original theory of Friedrich and Wintgen\cite{bicfw}.
In this study, we show that the branch at which the FW-BIC emerges also depends on the grating thickness.
On carefully inspecting the dispersion relation, we found that the position at which the BIC appears is fixed at the crossing of the original waveguide modes, while the coupled resonant modes of the entire system move owing to the change in the grating thickness, thus resulting in the flipping of the BIC branch.
We also show that the criteria discussed in our previous paper for the branch at which the BIC appears still hold.  
In addition, we observed that the magnitude of the anti-crossing gap, and hence, the internal (near field) coupling, varies with the change in the grating thickness.

In parallel, with the equation of motion of essentially the same form as that describes a system exhibiting the FW-BIC, the dynamics of non-Hermitian systems have been intensively investigated with an emphasis on the presence and the influence of exceptional points (EPs) at which the eigen solutions (resonant modes) formed from more than one oscillator with the mutual coupling coalesce\cite{PRA95_022117}.
In optics and photonics, EPs in active or passive parity-time (PT) symmetric structures as well as those in non-PT symmetric structures have attracted a significant amount of interest\cite{ep_rev1, ep_rev2} as such systems exhibit various exotic as well as practically important phenomena at or around the EP.
For example, 
asymmetric mode switching\cite{Nature537_76},
directional omni-polarizer\cite{PRL118_093002},
laser mode selection\cite{Science346_972, Science346_975,NatCom8_15389},
unidirectional invisibility or reflectionlessness\cite{NanoPhoto6_977},
directional total absorption\cite{OEX24_22219},
loss-induced transparency\cite{PRL103_093902},
polarization control\cite{PRB98_165418}, and
enhancement of Sagnac sensitivity\cite{Nature576_65, Nature576_70}
have been proposed and/or demonstrated.
They are expected to lead to a new paradigm of optical systems.

Though many of the reports on EPs have dealt with the case of no external (radiation) coupling, which is the origin of the FW-BIC, the EP is expected to appear even with the radiation coupling, and hence, the BIC and EP are closely located in the parameter space \cite{PRA95_022117, PRL123_023902}.
In the last part of this article, we show that the EP can indeed be realized near the BIC point in our device, where the tuning of the internal coupling with the grating thickness plays a significant role.
The continuous controllability of the system with the use of such an additional parameter is expected to encourage experimental investigations and the application of the BIC- and EP-related phenomena in optics and photonics\cite{PRL123_213903}.

\section{Device structure and the methods of analysis\label{Device structure}}

\begin{figure}[ht!]
\centering\includegraphics[width=5.5cm]{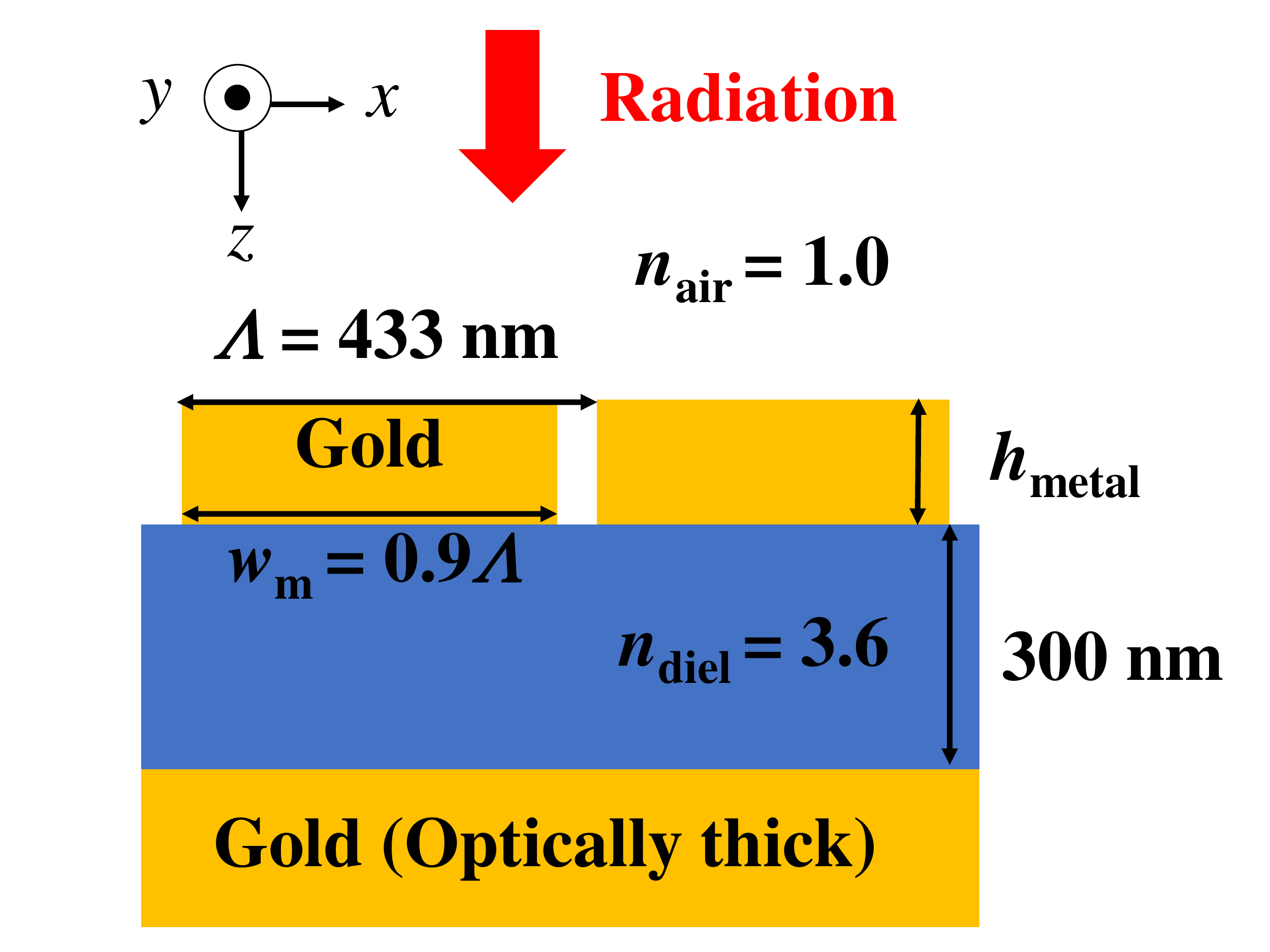}
\caption{Cross-sectional structure of the device considered in this work.}
\label{fig:system_cross_section}
\end{figure}

The system considered in this study is a dielectric waveguide sandwiched by a metal grating and optically thick backing metal, which is essentially the same structure as that considered in our previous paper \cite{KikkawaNJP}, as shown in Fig.\ \ref{fig:system_cross_section}. 
In this paper, we only vary the grating thickness $h_\mathrm{metal}$ and fix the other parameters.
We consider the region of the wavelength where only the 0th order wave is radiative in the air region.
For the permittivity of the gold, unless otherwise stated, the Drude-Lorentz model fitting the reported experimental data\cite{johnchristy} was used.
In some cases, however, we eliminated the imaginary part of the permittivity to neglect the loss and simplify the theoretical analysis.
We will specify the elimination when it is done.
We consider a case where a $P$-polarized wave is incident from the air region.
Owing to the backing metal, the incident light is totally reflected unless it is absorbed due to the ohmic loss at the metal surface.
Strong absorption occurs when the incident light excites the resonant mode, which comprises the waveguide modes in the dielectric region with a modification of the slit-induced internal and external (radiation) coupling.
The dominant waveguide modes for the situation considered in this paper are the first (lowest) and second transverse magnetic (TM) modes, $\mathrm{TM}_{0}$ and $\mathrm{TM}_{1}$, respectively, in the metal-dielectric-metal (MIM) structure.

For the analysis of the electromagnetic field and the resonant (eigen) mode in the device, a spatial coupled-mode method (CMM) \cite{cmm1} was used.
In the CMM, the wave propagation in the grating layer was described by that in a slit sandwitched by the nearby metal bars and the modes were calculated taking into account the permittivity of gold \cite{cmm2,cmm3}.
For the slit width of 43.3 nm considered here, only one propagating mode is present.
In addition, a temporal coupled-mode theory (TCMT) \cite{tcmt1, tcmt2} was used to analyze the response obtained by the CMM.
It should be noted that the CMM and TCMT are essentially different from each other; the former is a numerical solver of the EM field based on Maxwell's equations, while the latter is an equation that describes the motion of the oscillators corresponding, in the present case, to the original $\mathrm{TM}_{0}$ and $\mathrm{TM}_{1}$ modes with mutual internal and external coupling.

\section{Absorption spectra and the resonant mode of the system \label{Calculation results}}

\begin{figure}[ht!]
\centering\includegraphics[width=11cm]{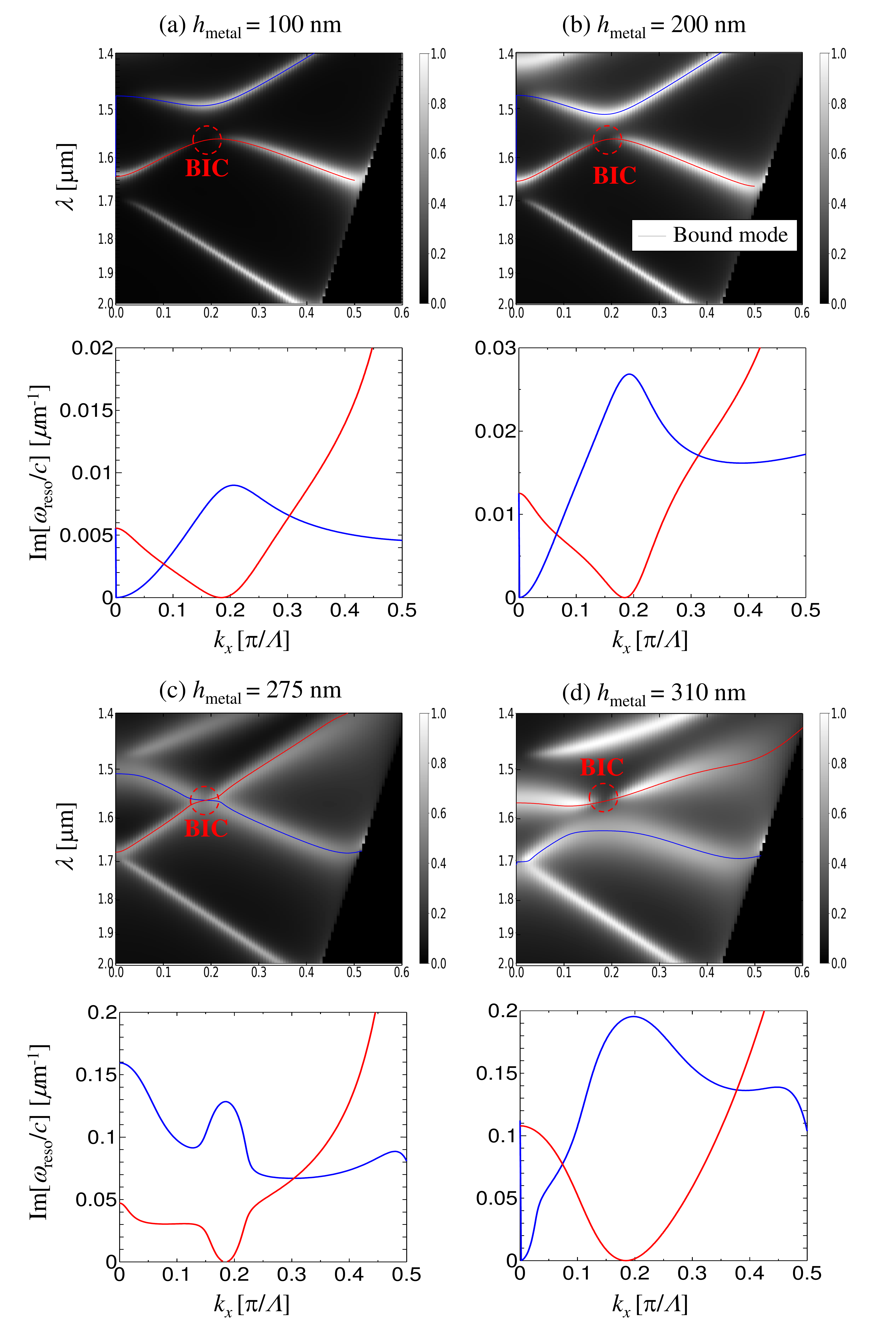}
\caption{Dispersion relation of the absorption and the resonant mode for the four cases of  $h_{\mathrm{metal}}$: (a) 100 nm, (b) 200 nm, (c) 275 nm, and (d) 310 nm. In the upper panel, the absorption is depicted by the gray scale along with the wavelength corresponding to the real part of the resonant-mode frequency indicated by the blue and red lines. In the lower panel, the imaginary part of the angular frequency of the resonant mode is plotted using the same color as that used for the same branch in the real part. The frequency was normalized by the speed of the light $c$  in vacuum.
It should be noted that the wavelength is given in the inverted scale (linear in frequency). The dashed circles indicate the position of the BIC.
}
\label{fig:dispersion_relation}
\end{figure}

Figure\ \ref{fig:dispersion_relation} shows the dispersion relation of the absorption for the wave incident from the air region in the gray-scale maps for the four cases of the grating thickness $h_\mathrm{metal}$.
The solid lines  on the absorption map indicate the wavelength corresponding to the real part of the resonant mode frequency.
It can be observed that the high-absorption bands appear along the resonant modes.
As mentioned above, each resonant mode can be assigned to an originating MIM mode, except in the case of the anti-crossing region wherein the modes are mixed by the slit.
In all the cases of the grating thickness, as indicated by the dashed circles, the absorption disappears locally near the anti-crossing.
This behavior is a signature of the emergence of the BIC.
The imaginary part of the resonant mode frequency is depicted in the lower panel for each $h_\mathrm{metal}$ with the same colors for each branch as those used for the real part.
In the calculation of the resonant modes, the imaginary part of the metal permittivity was set to zero such that the imaginary part of the resonant mode frequency is only due to the radiation loss.
The disappearance of the absorption corresponds to the vanishing imaginary part, which confirms that the disappeared mode is indeed the BIC. 

There are two remarkable features of the absorption map and resonant mode dispersion.
First, the magnitude of the anti-crossing gap depends on $h_\mathrm{metal}$.
With the increase in $h_\mathrm{metal}$, the gap first becomes small and almost diminishes for $h_\mathrm{metal}\sim275\ \mathrm{nm}$.
The gap then reappears with a further increase in $h_\mathrm{metal}$.
Second, the branch at which the BIC appears changes as  $h_\mathrm{metal}$ crosses $\sim275\ \mathrm{nm}$, where the gap shrinkage occurs.
The BIC appears on the lower branch for $h_\mathrm{metal}<275\ \mathrm{nm}$ and on the higher branch for $h_\mathrm{metal}>275\ \mathrm{nm}$.
In the following, we discuss these features in detail, in the order of 1) the position and branch of the BIC and 2) the variation of the anti-crossing gap in Sec.\ 4 and Sec.\ 5, respectively.

\section{Position of the BIC \label{Sec_BIC_position}}

\subsection{The position of the BIC in terms of the empty lattice mode\label{Sec_BIC_position_empty}}

\begin{figure}[ht!]
\centering\includegraphics[width=11cm]{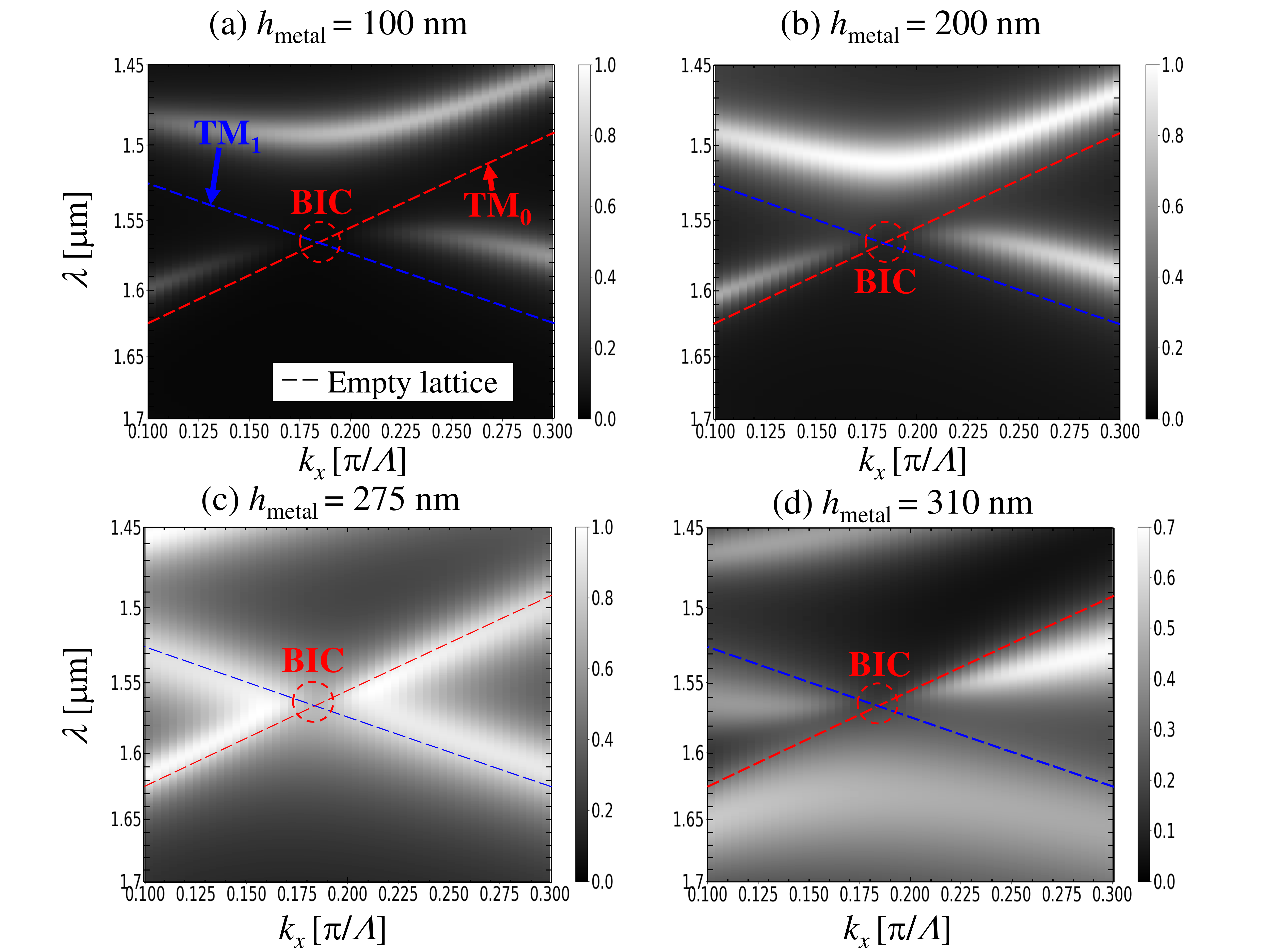}
\caption{Magnified view of the absorption map shown in Fig.\ \ref{fig:dispersion_relation}. The dashed circles and dashed lines indicate the position of the BIC and empty lattice bands, respectively}
\label{fig:bound_modes_cmm}
\end{figure}

Here, we first discuss the position and the branch of the BIC.
In Fig.\ \ref{fig:bound_modes_cmm}, we present a magnified view of Fig.\ \ref{fig:dispersion_relation} near the anti-crossing.
The position of the BIC is indicated by the dashed circles.
As mentioned above, the BIC appears on the lower (upper) branch for $h_\mathrm{metal} < 275\ (> 275)\ \mathrm{nm}$.
For $h_\mathrm{metal}=275\ \mathrm{nm}$, the anti-crossing disappears, and the BIC appears at the crossing point.
On observing the figure more closely, we notice that the position of the BIC is not moved by the change in $h_\mathrm{metal}$.
The broken lines plotted in the figures are the empty lattice $\mathrm{TM}_{0}$ and $\mathrm{TM}_{1}$ waveguide modes, which were obtained by folding the dispersion curves of these modes in the MIM waveguide with flat metals (no slit) into the first  Brillouin Zone.
Interestingly and importantly, it can be clearly observed that the BIC is always located at the crossing point of the two empty lattice bands, which does not move with the change in the grating thickness.
Therefore, the branch inversion of the BIC that is observed in Fig.\ \ref{fig:dispersion_relation} is understood to be a result of the movement of the leaky mode associated with the change in the grating thickness over the BIC point, which is fixed at the crossing of the empty lattice dispersion.

The appearance of the BIC at the crossing point of the empty lattice bands can be explained using the CMM as follows.
For the BIC solution, the radiative fields in the air region become null owing to the nature of the BIC.
It can be realized when the fields inside the slit are null, if the slit is single mode as in the present device.
In such cases, the fields in the dielectric region satisfy the same relation at the slit/waveguide interface as that of the empty lattice MIM modes, which means that the fields inside the dielectric region are those of the empty lattice modes when the BIC occurs.
It should be mentioned that, if the slit is not single mode, the null field in the air region does not mean the null field in the slit and the BIC point can be shifted from the crossing point of the empty lattice.

\begin{figure}[ht!]
\centering\includegraphics[width=11cm]{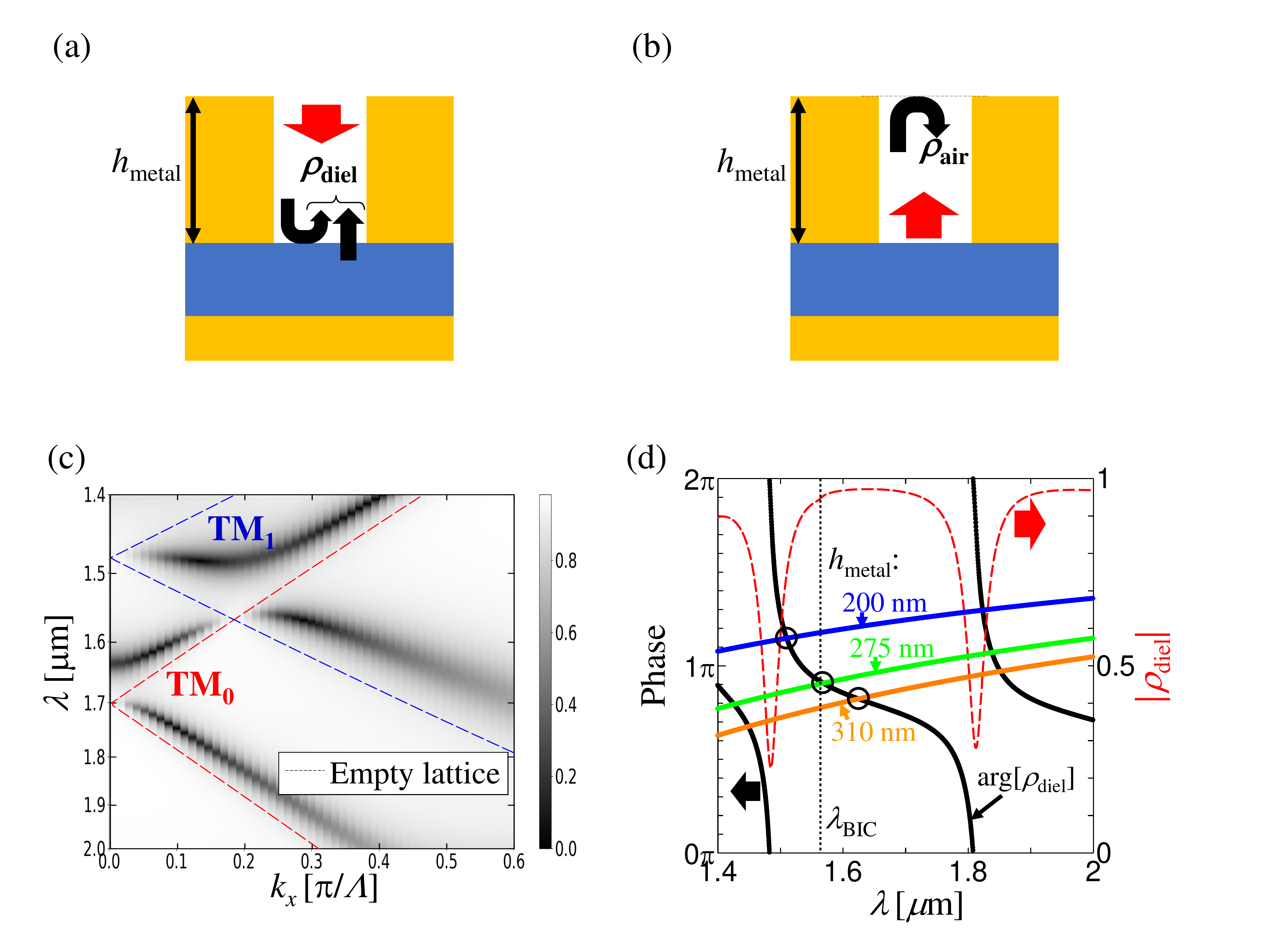}
\caption{(a) Schematic showing the reflection coefficient $\rho_{\mathrm{diel}}$ at the slit/dielectric interface for the slit mode.
(b) Schematic showing the reflection coefficient $\rho_{\mathrm{air}}$ at the slit/air interface for the slit mode. 
(c) $|\rho_{\mathrm{diel}}|$ as a function of $k_x$ and the wavelength depicted in gray scale. The empty lattice bands of the waveguide are superimposed by the dashed lines.
(d) Amplitude (black solid line) and phase (red dashed line) of $\rho_{\mathrm{diel}}$ at $k_{x} = 0.185[\pi/\mathit{\Lambda}]$ for which the BIC appears. The blue, green, and orange lines indicate the value of arg$(\rho_{\mathrm{diel}})$ required to satisfy Eq.\ \ref{fp_cond} for  the cases of $h_{\mathrm{metal}}$ = 200, 275, and 310 nm, respectively. The solid circle indicates the intersection of these lines and arg$(\rho_{\mathrm{diel}})$, which gives the FP resonance.
}
\label{fig:reflection at interface}
\end{figure}

As another method of understanding the system, we calculated the reflection coefficient $\rho_{\mathrm{diel}}$ for the wave approaching the slit/dielectric interface from the infinitely thick grating as schematically depicted in Fig.\ \ref{fig:reflection at interface} (a).
Figure \ref{fig:reflection at interface} (c) shows the dispersion of $|\rho_{\mathrm{diel}}|$ along with the empty lattice bands (indicated by dashed lines).
It can be observed that the anti-crossing of the bands and the BIC occurs and that the BIC is located at the crossing of the empty lattice band, which is not affected by the grating. 
Then let us elucidate the role of the finite thickness of the grating.
Indicated by the black solid line in Fig.\ \ref{fig:reflection at interface}(d) is $\mathrm{arg}(\rho_{\mathrm{diel}})$ as a function of the wavelength for $k_x=0.185$ [$\pi/\Lambda$], where the BIC emerges at $ \lambda= 1563 $ nm.
It should be noted that $\mathrm{arg}(\rho_{\mathrm{diel}})$ is affected only by the radiative mode because the BIC mode is not coupled to the slit as mentioned above.
The red dashed line is $|\rho_{\mathrm{diel}}|$.
Between the dips in $|\rho_{\mathrm{diel}}|$, which correspond to the coupled $\mathrm{TM}_0$-$\mathrm{TM}_1$ mode at approximately 1500 nm and uncoupled $\mathrm{TM}_0$ mode at approximately 1800 nm,  $\mathrm{arg}(\rho_{\mathrm{diel}})$ varies gradually from $ 2 \pi $ to 0. 
The wavelength of the BIC $ \mathit{\lambda}_{\mathrm{BIC}} $ is indicated by the vertical dotted line.
Now, let us hypothesize that the solution of the radiative mode of the entire system with a grating of finite thickness can be viewed as the Fabry-Perot (FP) resonance for the wave propagating in the slit, which is given by 
\begin{equation}
\label{fp_cond}
\mathrm{arg}(\rho_{\mathrm{air}})+\mathrm{arg}(\rho_{\mathrm{diel}})+2\beta_{\mathrm{slit}} h_{\mathrm{metal}} = 2 \pi m,
\end{equation}
where $\mathrm{arg}(\rho_{\mathrm{air}}$) is the phase of the reflection at the slit/air interface (See Fig.\ \ref{fig:reflection at interface}(b)), $\beta_{\mathrm{slit}}$ is the propagation constant in the slit, and $m$ is an integer.
The blue, green, and orange lines in Fig.\ \ref{fig:reflection at interface}(d) indicate the value of $\mathrm{arg}(\rho_{\mathrm{diel}})$ required to satisfy Eq.\ \ref{fp_cond} for $h_{\mathrm{metal}}$ = 200, 275, and 310 nm, respectively.
Hence, the point at which the line for each $h_{\mathrm{metal}}$ crosses $\mathrm{arg}(\rho_{\mathrm{diel}})$ (indicated by the black circles) provides the wavelength that satisfies Eq.\ \ref{fp_cond}.
For $h_{\mathrm{metal}}$ = 200 nm, the wavelength of the crossing is shorter than $ \mathit{\lambda}_{\mathrm{BIC}} $.
This corresponds to Fig.\ \ref{fig:bound_modes_cmm} (b) where the BIC is located on the longer-wavelength branch.
For $h_{\mathrm{metal}}$ = 275 nm, the crossing is located near $ \mathit{\lambda}_{\mathrm{BIC}} $, which corresponds to Fig.\ \ref{fig:bound_modes_cmm} (c).
Then, for $h_{\mathrm{metal}}$ = 310 nm, the wavelength of the crossing is greater than $ \mathit{\lambda}_{\mathrm{BIC}} $, which corresponds to Fig.\ \ref{fig:bound_modes_cmm} (d).
It can be observed that the wavelength of the radiative mode in Fig.\ \ref{fig:bound_modes_cmm} at $k_x=0.185$ [$\pi/\Lambda$], where the BIC emerges, is predicted very well, which justifies the hypothesis, and that the thickness $h_{\mathrm{metal}}$ of the grating can be said to determine the wavelength that satisfies the FP condition in the slit for the radiative (non-BIC) solution.
It should be mentioned again that the discussions presented here are valid for the case in which the wave propagation in the slit is in the single mode or only one mode is dominating.

\subsection{Temporal coupled-mode theory appropriate for the present system \label{Sec_TCMT}}

The response of the present device can be analyzed using TCMT \cite{KikkawaNJP} for the two resonators corresponding to the empty-lattice TM$_0$ and TM$_1$ modes with the internal and external coupling.
The time evolution of these two resonators $\mathbf{a}^\mathrm{T}=(a_{1} , a_{2})$ driven by an external field $s_{+}$ with a coupling coefficient $\mathbf{D}$ can be written as
\begin{equation}
\label{tcmt_formula}
\frac{d}{dt}\mathbf{a}=-i\{\mathbf{\Omega}-i(\mathbf{\Gamma}_{i}+\mathbf{\Gamma}_{e})\}\mathbf{a}+\mathbf{D}^\mathrm{T}s_{+},
\end{equation}
where the  matrices $\mathbf{\Omega}$, $\mathbf{\Gamma}_{i}$, and $\mathbf{\Gamma}_{e}$ represent the eigenfrequencies with internal coupling, internal loss, and radiation loss, respectively.
Though the framework is essentially the same as that in a previous study\cite{KikkawaNJP},  to analyze the present system, it is necessary to modify $\mathbf{\Omega}$ in order to keep the BIC solutions fixed at the degenerate point of the two basis modes.
As presented below, the matrix $\mathbf{\Omega}$ that satisfies this requirement is,
\begin{equation}
\label{terms}
\mathbf{\Omega}=\left(\begin{array}{@{\,}cccc@{\,}}\tilde{\omega}_{1}\equiv\omega_{1}+p\alpha\sqrt{\frac{\gamma_{e1}}{\gamma_{e2}}} & \alpha \\ \alpha & \tilde{\omega}_{2}\equiv\omega_{2}+p\alpha\sqrt{\frac{\gamma_{e2}}{\gamma_{e1}}} \\\end{array}\right),
\end{equation}
where $\omega_{1,2}$ and $\gamma_{e1,2}$ are the eigenfrequency and external (radiation) loss of modes 1 or 2, respectively; $\alpha$  is the internal coupling; and $p$ is the phase difference between the two modes for coupling to the external field.
Here, $p$ has the value of either +1 or $-$1; $p=+1$ if the two modes are in phase (no phase difference), and $p=-1$ if the two modes are out of phase ($\pi$ phase difference)\cite{KikkawaNJP}. 
The modification comprises the addition of $ p\alpha\sqrt{\frac{\gamma_{e1}}{\gamma_{e2}}} $ and $ p\alpha\sqrt{\frac{\gamma_{e2}}{\gamma_{e1}}} $ to the diagonal terms of $ \mathbf{\Omega} $.
The matrices
\begin{equation}
\mathbf{\Gamma}_{e}=\left(\begin{array}{@{\,}cccc@{\,}}\gamma_{e1} & \gamma_0  \\ \gamma_0 & \gamma_{e2} \\\end{array}\right),
\quad
\mathbf{D}^\mathrm{T}=e^{i\varphi_{d}}\left(\begin{array}{@{\,}cc@{\,}} \sqrt{2 \gamma_{e1}}  \\  p\sqrt{2 \gamma_{e2}} \\\end{array}\right),
\end{equation}
where $\gamma_0 = p\sqrt{\mathbf{\gamma}_{e1}\mathbf{\gamma}_{e2}}$ represents the radiation coupling and $ \varphi_{d} $ is an arbitrary phase, are the same as before, and the matrix
\begin{equation}
\mathbf{\Gamma}_{i}=\left(\begin{array}{@{\,}cccc@{\,}} \gamma_{i1} & 0 \\ 0 & \gamma_{i2} \\\end{array}\right),
\end{equation}
where  $\gamma_{i1,2}$ is the internal loss (in the metal in the present device) of each mode, was included in Eq.\ \ref{tcmt_formula} for generalization. 
The expression for $ \mathbf{D} $ and the off-diagonal terms of $ \mathbf{\Gamma}_{e} $  are derived using the time-reversal symmetry and energy conservation law\cite{tcmt2}.

For the time dependence of $ e^{-i\omega t} $, the eigenfrequencies of the resonant modes are obtained \cite{bicfw, PRA95_022117, PRC67_054322, PLA145_265} from Eq.\ \ref{tcmt_formula} as the solution of
\begin{eqnarray}
\label{eigen_eq}
\{\omega-(\tilde{\omega}_{1}-i\gamma_{t1})\} \{\omega-(\tilde{\omega}_{2}-i\gamma_{t2})\} +(ip\alpha+\sqrt{\gamma_{e1}\gamma_{e2}})^2 =0
\end{eqnarray}
where $\gamma_{t1, 2}=\gamma_{i1, 2}+\gamma_{e1, 2}$.
The condition for one of the solutions being the BIC can be derived from Eq.\ \ref{eigen_eq}, by assuming that one of the solutions is purely real\cite{KikkawaNJP}.
On assuming $\gamma_{i1}=\gamma_{i2}=0 $ for simplicity, we obtain the condition
\begin{equation}
\label{bic_cond}
\omega_{1}=\omega_{2} \equiv \omega_{0}.
\end{equation}
Then, the solutions are
\begin{equation}
\label{bic_solution}
\omega=
\left\{
\begin{array}{l}
\omega_{0}+p\alpha(\sqrt{\frac{\gamma_{e1}}{\gamma_{e2}}}+\sqrt{\frac{\gamma_{e2}}{\gamma_{e1}}})-i(\gamma_{e1}+\gamma_{e2}) \\ 
\omega_{0}
\end{array}
\right.
.
\end{equation}
The first solution is radiative (leaky) with the loss $\gamma_{e1}+\gamma_{e2}$, and the second solution is the BIC, which has no imaginary part.
Thus, the formulation is confirmed to provide the BIC that always lies at the crossing point of the original eigenfrequencies, $\omega_{0}=\omega_{1}=\omega_{2}$.

\subsection{Prediction of BIC-branch in terms of $p\alpha$ \label{pa_predict}}

In the preceding subsections, we showed that the BIC appears at the crossing point of the empty-lattice modes. 
In our previous report\cite{KikkawaNJP}, we showed that the branch at which the BIC appears is determined by the sign of $p\alpha$.
Herein, we show that the aforementioned rule still holds.
In the TCMT analysis, as observed in Eq.\ \ref{bic_solution}, the branches at which the BIC appears are determined as follows:
\begin{equation}
\label{bic_branch}
p\alpha \ \ 
\left\{
\begin{array}{l}
>0 \quad\mbox{(lower branch)}\\ 
<0 \quad\mbox{(upper branch)}
\end{array}
\right. ,
\end{equation}
as in the case without modification\cite{KikkawaNJP}. 
Therefore, we check sign$(p\alpha)$ for the cases of $h_\mathrm{metal} < 275\ \mathrm{nm}$ and $h_\mathrm{metal} > 275\ \mathrm{nm}$, where the BIC appears at the upper and lower branches, respectively, as shown in Fig.\ \ref{fig:dispersion_relation} or Fig.\ \ref{fig:bound_modes_cmm}.

\begin{figure}[ht!]
\centering\includegraphics[width=11cm]{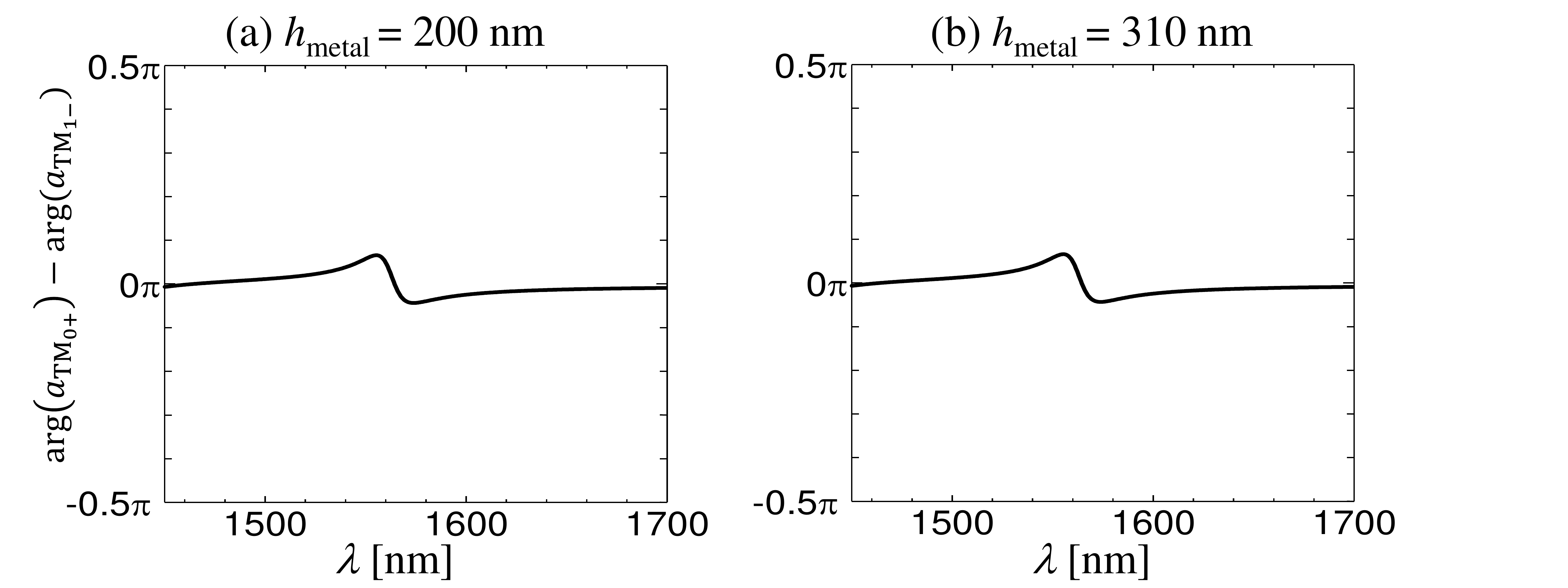}
\caption{Phase difference between the amplitudes of the two waveguide modes near the anti-crossing ($k_x=0.1839 [\pi/\mathit{\Lambda}]$) as a function of wavelength. (a) $h_\mathrm{metal}=200\ \mathrm{nm}$ and (b) $h_\mathrm{metal}=310\ \mathrm{nm}$.}
\label{fig:p_value}
\end{figure}

The evaluation of sign$\{p\alpha\}$ was conducted in the same manner as before\cite{KikkawaNJP}, i.\ e.\, by expanding the electromagnetic fields inside the dielectric waveguide of the empty-lattice waveguide modes.
In the present case, the dominant modes are the right-propagating TM$_0$ and the left propagating TM$_1$ modes near the anti-crossing point. 
Sign$(p)$ and  sign$(\alpha)$ are then evaluated from the complex amplitude of each mode obtained in the expansion.
Sign$(p)$ is evaluated directly from the phase difference between the complex amplitudes.
Figure \ref{fig:p_value}(a) and (b) show the calculated phase difference at $k_{x} = 0.1839[\pi/\mathit{\Lambda}]$, which is slightly displaced from the point at which the BIC emerges, for $h_\mathrm{metal}=$ 200 nm and 310 nm, respectively.
At approximately $\mathit{\lambda} = 1550 \mathrm{nm}$, the phase is disturbed because the coupled resonance is excited.
However, the phase converges to 0 far away from the resonant wavelength in both cases.
Therefore, it can be concluded that $p=1$ in both cases.

Sign$(\alpha)$ is evaluated based on the phase difference between the amplitudes at the two solutions  of the coupled resonant mode with the upper ($\omega_{+}$) and lower ($\omega_{-}$) frequencies, as the coupled mode oscillation is viewed as the bonding or anti-bonding solution depending on the sign of the coupling constant $\alpha$.
The phase difference for $h_\mathrm{metal}=200\ \mathrm{nm}$ and $310\ \mathrm{nm}$ are summarized in Table.\ \ref{tab:table_alpha}. For $h_\mathrm{metal}=200$ nm, the two modes are in phase (out of phase) at $\omega_{+}$ ($\omega_{-})$, which corresponds to $\alpha>0$,
while for $h_\mathrm{metal}=310$ nm, the two modes are out of phase (in phase) at $\omega_{+}$ ($\omega_{-}$), which corresponds to $\alpha<0$.
Therefore, we can conclude that $p\alpha$ is positive for $h_\mathrm{metal}=200\ \mathrm{nm}$ and negative for $h_\mathrm{metal}=310\ \mathrm{nm}$, which confirms that the criteria of Eq.\ \ref{bic_branch} are satisfied.
We can also say that the sign$(p\alpha)$ is controlled by the grating thickness.

\begin{table}[ht!]
\caption{Phase difference between the amplitudes of the two waveguide modes near the anti-crossing ($k_x=0.1839 [\pi/\mathit{\Lambda}]$) at the two coupled resonant modes $\omega_{+}$ and $\omega_{-}$.  (a) $h_\mathrm{metal}=200\ \mathrm{nm}$ and (b) $h_\mathrm{metal}=310\ \mathrm{nm}$.}
\centering\includegraphics[width=11cm]{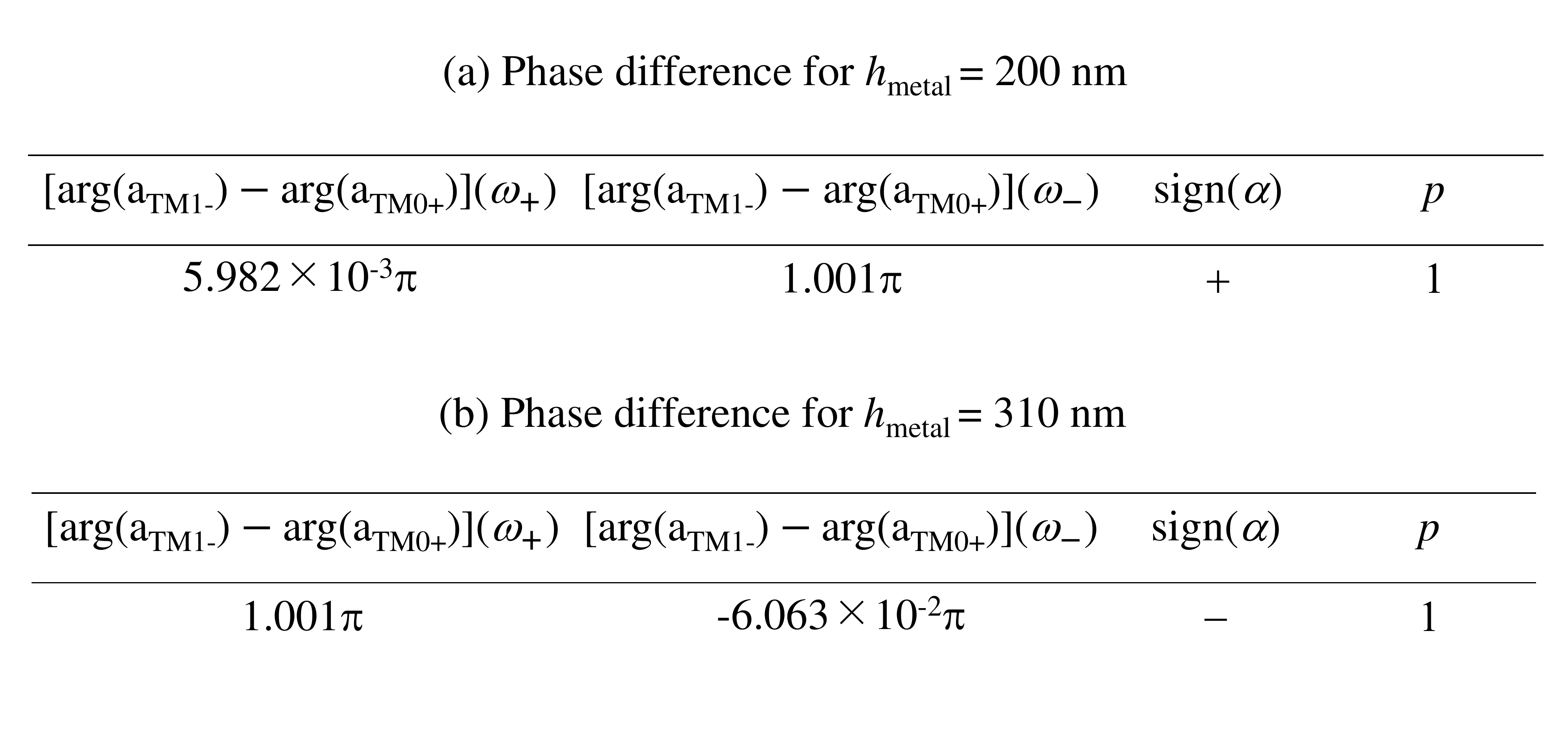}
\label{tab:table_alpha}
\end{table}

\section{Variation of the anti-crossing gap and the emergence of exceptional point \label{Sec_EP}}
\subsection{Grating thickness dependence of the anti-crossing gap \label{Sec_alpha value}}

\begin{figure}[ht!]
\centering\includegraphics[width=5.5cm]{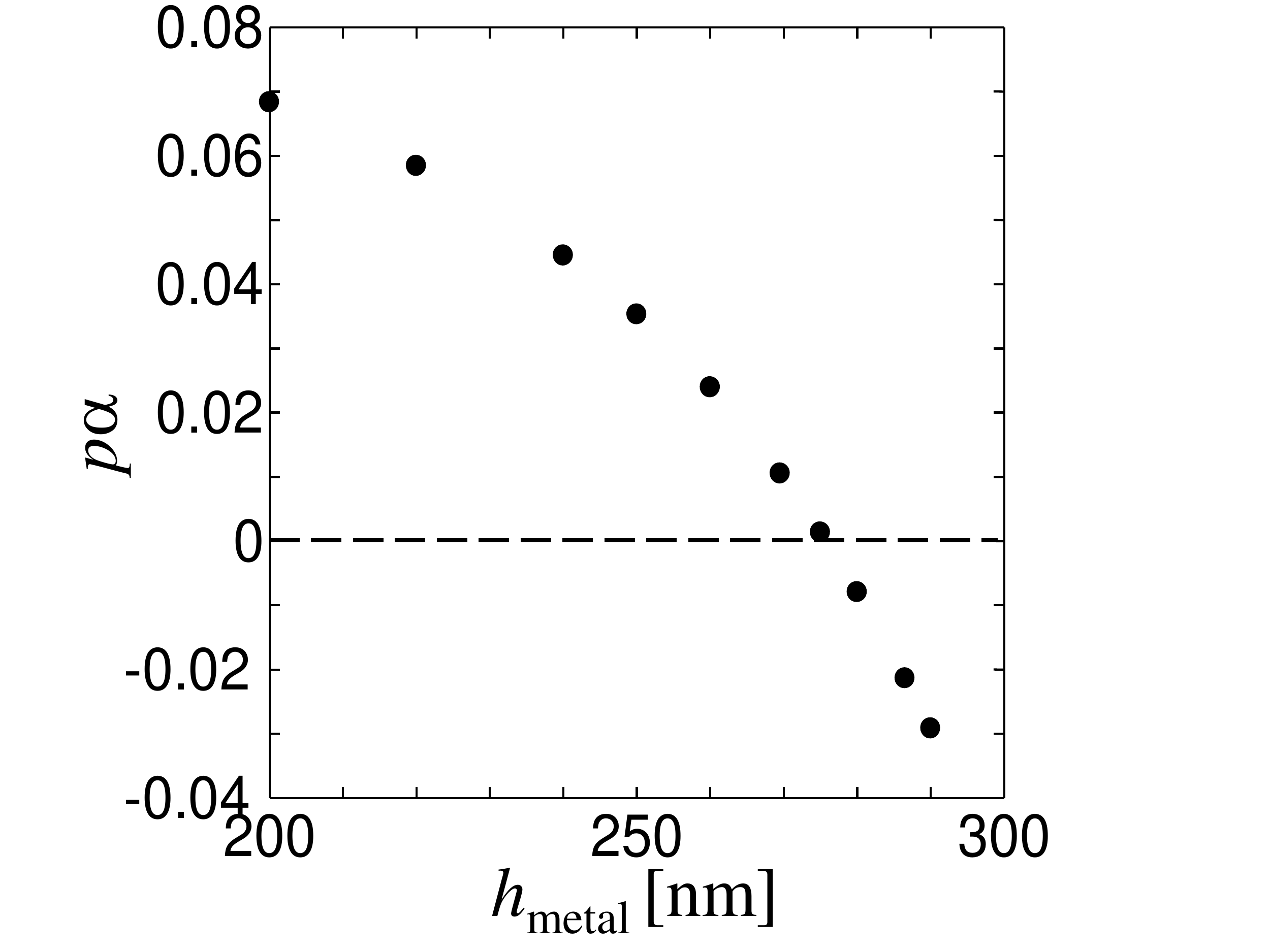}
\caption{Value of $p\alpha$ as a function of  the grating thickness calculated using Eq.\ $\ref{differ_bic}$. Here $p=1$ and the sign$(\alpha)$ were selected in accordance with the result obtained in Sec.\ 4.3.}
\label{fig:abs_pa}
\end{figure}

In this section, we discuss the second feature of the device, namely the variation of the anti-crossing gap with the change of the grating thickness, $h_{\mathrm{metal}}$.
As seen in Eq.\ \ref{bic_solution}, the difference in the real part of the angular frequency $\mathrm{Re}\{\Delta\omega\}$ between the non-BIC (radiating) and the BIC solutions is 
\begin{equation}
\label{differ_bic}
\mathrm{Re}\{\Delta\omega\} = \frac{p\alpha}{2}(\sqrt{\frac{\gamma_{e1}}{\gamma_{e2}}}+\sqrt{\frac{\gamma_{e2}}{\gamma_{e1}}}).
\end{equation}
Here, as mentioned above, $p=1$ and sign$(\alpha)$ depends on $h_{\mathrm{metal}}$.
Although the radiation losses, $\gamma_{e1}$ and $\gamma_{e2}$, are not exactly equal in our device, they are not much different, which can be observed from the linewidth of the two bright lines in Fig.\ \ref{fig:dispersion_relation}.
Therefore, we evaluate the value $p\alpha$ using an approximated relation $p\alpha=\Delta\omega$, where $\Delta\omega$ is evaluated based on the real part of the resonant-mode frequencies at $k_x=0.185$ [$\pi/\Lambda$], where the BIC appears. 
The result is shown in Fig.\ \ref{fig:abs_pa}, where sign$(\alpha)$ was selected according to the result presented in Sec.\ \ref{pa_predict}. 
The value of $p\alpha$ decreases as $h_{\mathrm{metal}}$ increases and crosses zero at approximately $h_\mathrm{metal}=275\ \mathrm{nm}$.

\subsection{Emergence of the exceptional point \label{Sec_emergence_EP}}

Recently, with the same type of equation of motion as the TCMT (Eq.\ \ref{tcmt_formula}) used above, the physics and the application of EPs in optics and photonics have been intensively investigated \cite{ep_rev1, ep_rev2}.
Hence, it is interesting to consider the possibility of the realization of EP in the device considered here. 
The eigenfrequencies of the system are obtained \cite{PRA95_022117, PRC67_054322} as the solution of Eq.\ \ref{eigen_eq}.
\begin{equation}
\label{eigen_solutions}
\omega = \frac{ (\tilde{\omega}_{1}+\tilde{\omega}_{2})-i(\gamma_{t1}+\gamma_{t2} ) \pm \sqrt{D} }{2},
\end{equation}
where
\begin{equation}
\label{sqrt_of_eigen_solutions}
D \equiv \{(\tilde{\omega}_{1}-\tilde{\omega}_{2})^2-(\gamma_{t1}-\gamma_{t2})^2+4(\alpha^2-\gamma_{0}^2) \} 
-i 2\{ (\tilde{\omega}_{1}-\tilde{\omega}_{2})(\gamma_{t1}-\gamma_{t2})+ 4 \alpha \gamma_{0} \}.
\end{equation}
The EP, where both the real and imaginary parts of the eigenfrequency coalesce, is realized when $D$ becomes zero, i.e., $ \mathrm{Re} \{D\}= \mathrm{Im} \{D\}=0 $.
$ \mathrm{Im} \{D\}=0 $ presents the following condition.
\begin{equation}
\label{im_d_zero_cond}
\omega_{1}-\omega_{2} = - \frac{ 4\alpha\gamma_{0} }{ \gamma_{t1}-\gamma_{t2} } -p\alpha\left(\sqrt{\frac{\gamma_{e1}}{\gamma_{e2}}}+\sqrt{\frac{\gamma_{e2}}{\gamma_{e1}}}\right),
\end{equation}
for which
\begin{equation}
\label{d_of_re_d_zero}
\mathrm{Re} \{D\}
  =\frac{ \{ (\gamma_{t1} - \gamma_{t2})^2 + 4 \gamma_{0}^2 \} \{ 4 \alpha^2 - (\gamma_{t1} - \gamma_{t2})^2 \} }{ (\gamma_{t1} - \gamma_{t2})^2 }
\end{equation}
becomes zero when $ 4 \alpha^2 = (\gamma_{t1} - \gamma_{t2})^2 $. 
The condition for the EP is then expressed as
\begin{equation}
\label{im_d_ep_cond1}
|\gamma_{t1} - \gamma_{t2}| = 2 |\alpha|,
\end{equation}
\begin{equation}
\label{im_d_ep_cond2}
\omega_{1}-\omega_{2} = \pm 2\gamma_{0} -p\alpha(\sqrt{\frac{\gamma_{e1}}{\gamma_{e2}}}+\sqrt{\frac{\gamma_{e2}}{\gamma_{e1}}})
=p\sqrt{\gamma_{e1}\gamma_{e2}} \left\{ \pm2-\alpha\left( \frac{1}{\gamma_{e2}}-\frac{1}{\gamma_{e1}}\right) \right\}.
\end{equation}
In our device, $\omega_1-\omega_2$ in Eqs. \ref{im_d_zero_cond} or \ref{im_d_ep_cond2} are selected based on $ k_{x} $.
The effect of the radiation loss $\gamma_{e1,2}$ and the radiation coupling $\gamma_0$ appears only in the form of the shift of the conditions.

\begin{figure}[ht!]
\centering\includegraphics[width=11cm]{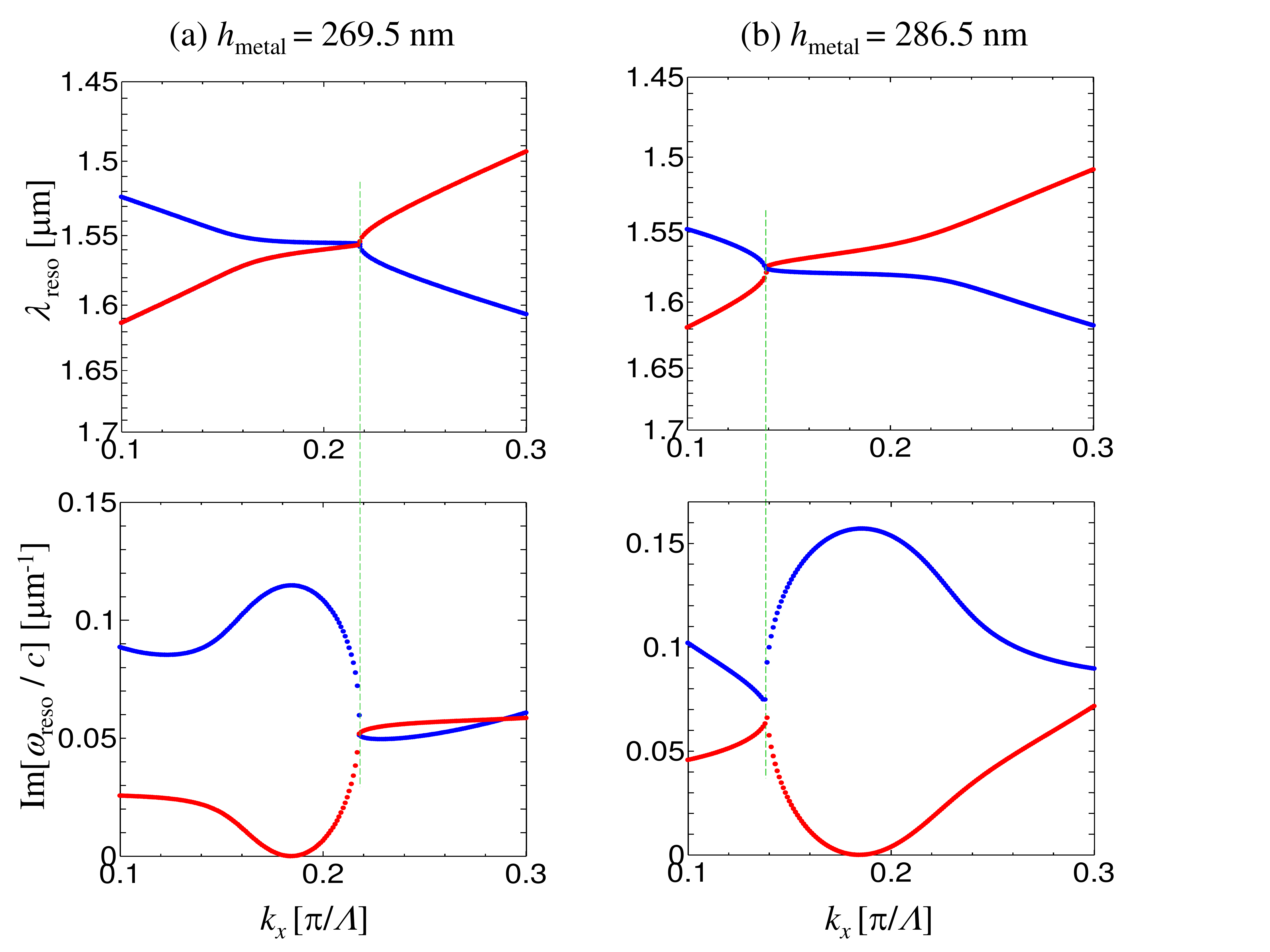}
\caption{Real (upper) and imaginary (lower) parts of the coupled resonant modes  for (a)$h_\mathrm{metal}=269.5\ \mathrm{nm}$ and (b)$h_\mathrm{metal}=286.5\ \mathrm{nm}$. The real part is given by the corresponding wavelength . For each $h_\mathrm{metal}$, there exists a point at which the real and imaginary parts coalesce, which indicates the EP.}
\label{fig:EP}
\end{figure}

Let us neglect the internal loss for the purpose of simplicity, i.\ e.\, $\gamma_{t1,2}=\gamma_{e1,2}$.
Although it is difficult to evaluate precisely the values of $\gamma_{e1,2}$ in the specific structure, $|\alpha|$ can be tuned by changing the grating thickness $h_{\mathrm{metal}}$ as shown in Fig.\ \ref{fig:abs_pa}.
Therefore, we can identify a case where the EP emerges on varying $h_{\mathrm{metal}}$.
In Fig.\ \ref{fig:EP}, we show the dispersion relation of the resonant mode frequency for $ h_{\mathrm{metal}} = 269.5$ nm and 286.5 nm, for which the coalescence of both the real and imaginary parts are observed at $ k_{x} = 0.217[\pi/\mathit{\Lambda}] $ and $ k_{x} = 0.140[\pi/\mathit{\Lambda}] $, respectively, thus demonstrating that the EPs are generated as expected.
Although not demonstrated here, the EP does not appear for other $h_{\mathrm{metal}}$.
For the two cases of $h_{\mathrm{metal}}$ for which the EP is realized, $\alpha$ is evaluated based on Fig.\ \ref{fig:abs_pa}, and the values of $\gamma_{e1,2}$ are then estimated by the fitting of the $k_{x}$-dependent eigenfrequencies of the TCMT to those of the CMM.
The results are $\alpha/c=0.01\ [1/\mu\mathrm{m}]$, $\gamma_{e1}/c=0.05\ [1/\mu\mathrm{m}]$, and $\gamma_{e2}/c=0.07\ [1/\mu\mathrm{m}]$ for $ h_{\mathrm{metal}} = 269.5$ nm and  $\alpha=-0.02/c\ [1/\mu\mathrm{m}]$, $\gamma_{e1}/c=0.06\ [1/\mu\mathrm{m}]$, and $\gamma_{e2}/c=0.1\ [1/\mu\mathrm{m}]$ for $ h_{\mathrm{metal}} = 286.5$ nm, where $c$ is the speed of light in vacuum.
Therefore, in the present case, Eq.\ \ref{im_d_ep_cond2} is approximated as

\begin{equation}
\label{im_d_ep_cond2_approx}
\omega_{1}-\omega_{2} \simeq \pm 2\gamma_{0},
\end{equation}
which causes the EPs to appear at $ k_{x}$ in a manner nearly symmetrical with respect to the crossing point of the empty lattice modes, and hence, the BIC point, which is consistent with Fig.\ \ref{fig:EP}.

Eq.\ \ref{d_of_re_d_zero} suggests that the eigenfrequencies have a split real part and degenerated imaginary part for $4\alpha^2 > (\gamma_{t1}-\gamma_{t2})^2$, which corresponds to the effective (passive) PT symmetric case and a split imaginary part and degenerated real part for $4\alpha^2 < (\gamma_{t1}-\gamma_{t2})^2$, which corresponds to the broken (effective) PT case\cite{PRL123_213903}, provided that Eq.\  \ref{im_d_zero_cond} is satisfied.
For the present device, we can observe the latter case for 269.5 nm $< h_{\mathrm{metal}} <$ 286.5 nm at a point between 0.217 $[\pi/\mathit{\Lambda}] > k_{x} >  0.140\ [\pi/\mathit{\Lambda}] $ and the former case outside these regions.
Moreover, the structure presented here is suitable for the introduction of the gain as the pumping can be performed either optically through the slit or electrically through the grating and back metal.
In either case of with or without gain, the tunability of the parameter $\alpha$ in our structure is expected to be useful in the control of the device.
Lastly, we highlight that the control of $ \alpha $, which is realized in this study by controlling the grating thickness, can be realized using non-built-in methods.
As discussed in Sec.\ 4.1, the dependence of the resonant mode of the entire system, and hence, $\alpha$, on the grating thickness can be attributed to the change in the FP resonance in the slit.
Therefore, we can expect that $ \alpha $ can be varied by filling the slit with a dielectric or applying a voltage between the neighboring metal bars of the grating when the slit is filled with an electro-optical material.
The latter method, if realized, would be particularly useful.

\section{Conclusion \label{Conclusion}}
In this paper, we reported the emergence of the FW-BIC and EP in a dielectric waveguide comprising a metal grating, while focusing on their dependence on the grating thickness.
For any grating thickness, the BIC emerges at one of the branches near the anti-crossing formed from the two waveguide modes TM$_0$ and TM$_1$ with internal (near field) and external (radiation) coupling via the slit of the grating.
It was determined that, with a change in the thickness, the coupled modes move with the varying anti-crossing gap.
The gap diminishes at a certain thickness, and the branch at which the BIC appears flips.
The change in the anti-crossing gap corresponds to the change in the internal coupling constant.

We showed that, when the slit is narrow to support single-mode propagation, the branch and position of the FW-BIC is determined by a simple rule: the FW-BIC appears at the crossing point between the two waveguide modes in the empty-lattice (zero slit-width) limit.
In addition, these results are consistent with the criteria for the branch at which the BIC appears based on the phase of each basis mode presented in our previous paper.
Owing to the dependence of the internal coupling on the grating thickness, we can find the cases in which the EP appears in the same device based only on the selection of the grating thickness, consistently with the prediction.
As the dependence of the anti-crossing gap on the grating thickness can be understood in terms of the FP resonance in the slit, tuning could be performed using other methods such as the voltage applied to the metal grating with the slit filled by an electro-optical material.
The BIC and EP in  the dielectric waveguide comprising a metal grating, particularly with such tunability, are expected to result in the development of functional and high-performance optical and photonic devices as well as to become a platform for the fundamental research of non-Hermitian systems.

\section*{Funding}
This work was supported by JSPS KAKENHI Grant Numbers JP18K04979 and JP18K04980.

\end{document}